\documentclass[aps,prb,reprint,groupedaddress]{revtex4-2}

\usepackage{amsmath,amssymb}
\usepackage{graphics}
\usepackage{epsfig}
\usepackage{color}
\usepackage{lineno}
\usepackage{graphicx}
\usepackage{epstopdf}

\begin{document}

\title{Pressure-induced superconductivity in topological insulator Ge$_2$Bi$_2$Te$_5$ and the evolution with Mn doping}
\author{Shangjie Tian$^{1,2,3,\dag}$, Qi Wang$^{4,5,\dag}$, Yuqing Cao$^{2,3}$, Ying Ma$^{6}$, Xiao Zhang$^{6,*}$, Yanpeng Qi$^{4,5,7,*}$, Hechang Lei$^{2,3,*}$, and Shouguo Wang$^{1,*}$}
\affiliation{
$^{1}$Anhui Provincial Key Laboratory of Magnetic Functional Materials and Devices, School of Materials Science and Engineering, Anhui University, Hefei 230601, China\\
$^{2}$School of Physics and Beijing Key Laboratory of Optoelectronic Functional Materials $\&$ MicroNano Devices, Renmin University of China, Beijing 100872, China\\
$^{3}$Key Laboratory of Quantum State Construction and Manipulation (Ministry of Education), Renmin University of China, Beijing 100872, China\\
$^{4}$State Key Laboratory of Quantum Functional Materials, School of Physical Science and Technology, ShanghaiTech University, Shanghai 201210, China\\
$^{5}$ShanghaiTech Laboratory for Topological Physics, ShanghaiTech University, Shanghai 201210, China\\
$^{6}$State Key Laboratory of Information Photonics and Optical Communications $\&$ School of Physical Science and Technology, Beijing University of Posts and Telecommunications, Beijing 100876, China\\
$^{7}$Shanghai Key Laboratory of High-resolution Electron Microscopy, ShanghaiTech University, Shanghai 201210, China
}
\date{\today}

\begin{abstract}
Introducing superconductivity  (SC) or magnetism into topological insulators (TIs) can give rise to novel quantum states and exotic physical phenomena.
Here, we report a high-pressure transport study on the TI Ge$_2$Bi$_2$Te$_5$ and its Mn-doped counterparts. The application of pressure induces a SC in Ge$_2$Bi$_2$Te$_5$, which shows a dome-shape phase diagram with the maximum $T_c$ of 7.6 K at 23 GPa. 
Doping Mn into Ge$_2$Bi$_2$Te$_5$ introduces an antiferromagnetic order at ambient pressure and strongly weakens the pressure-induced SC, demonstrating that magnetism and SC compete in this material system. 
Present study provides a new platform for investigating the interplay among band topology, magnetism, and SC.
\end{abstract}


\maketitle

\begin{figure*}
	\includegraphics[scale=0.18]{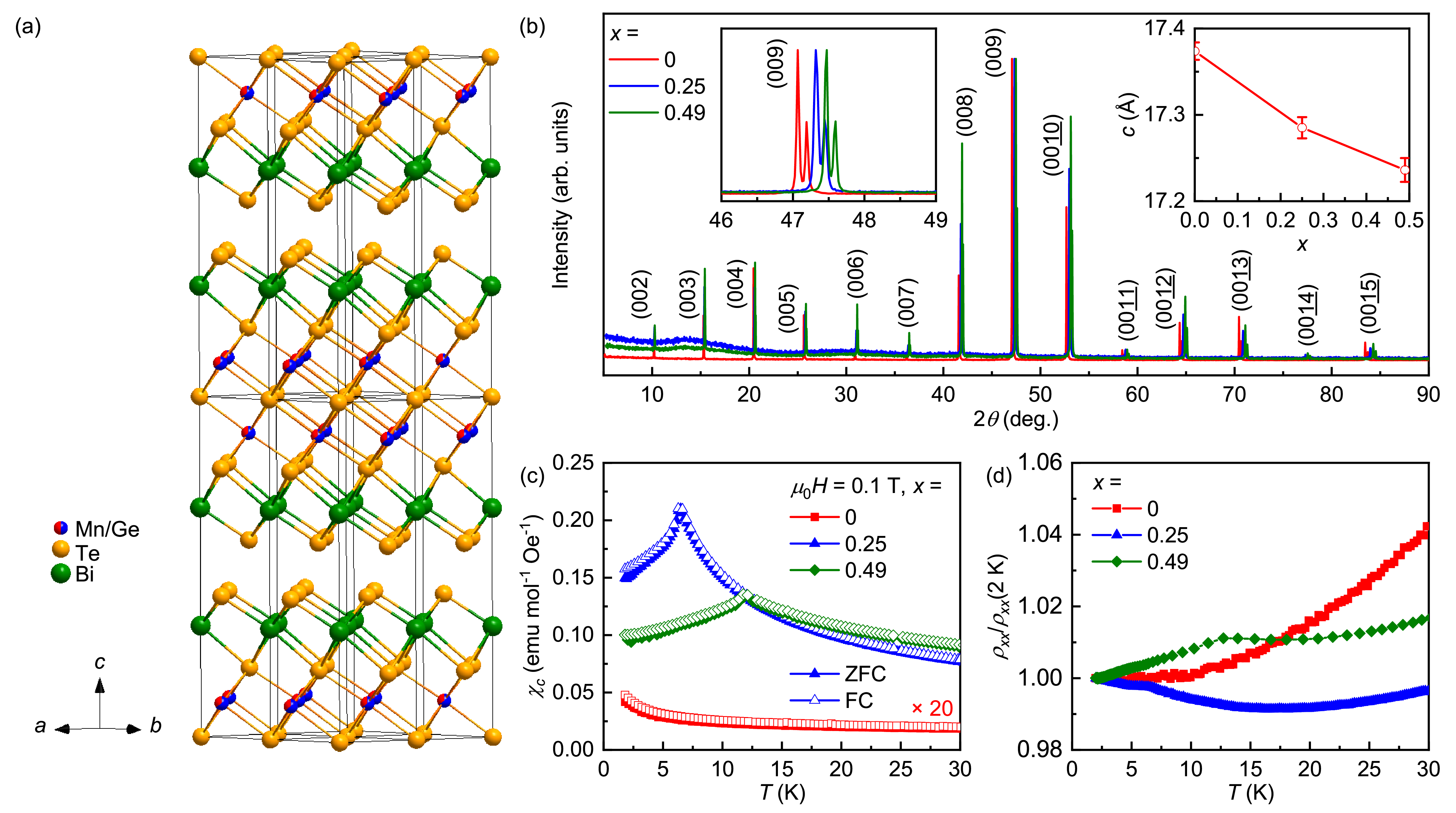}
	\caption{
		(a) Structure of (Ge$_{1-x}$Mn$_x$)Bi$_2$Te$_5$ single crystal. The red, blue, green and orange balls represent Ge, Mn, Bi and Te atoms, respectively. 
		(b) Single-crystal XRD patterns of (Ge$_{1-x}$Mn$_x$)$_2$Bi$_2$Te$_5$ single crystals. The left inset shows the enlarged part of (009) peaks. The right inset presents the fitted $c$-axial lattice parameter as a function of $x$.
		(c) Temperature dependence of $\chi_{c}(T)$ for (Ge$_{1-x}$Mn$_x$)$_2$Bi$_2$Te$_5$ single crystals under 0.1 T with ZFC and FC modes. The curves of Ge$_2$Bi$_2$Te$_5$ are magnified by a factor of 20 for clarity.
		(d) Normalized resistivity $\rho_{xx}(T)/\rho_{xx}$(2 K) as a function of temperature for (Ge$_{1-x}$Mn$_x$)$_2$Bi$_2$Te$_5$ single crystals.
	}
	\label{FigStr}
\end{figure*}

\section{Introduction}

Following the theoretical prediction and experimental confirmation of topological insulators (TIs), significant research interest has expanded to a broad class of materials exhibiting nontrivial band topology \cite{Fu2007,Hasa2010,Qi2011,Yan2012,Ando2013}. 
TIs are characterized by an insulating bulk state with robust, gapless surface states (SSs) protected by time-reversal symmetry \cite{Fu2007,Hasa2010,Qi2011}. 
The interplay of topology with magnetism or superconductivity (SC) leads to even richer exotic phenomena. 
For instance, when magnetism is introduced into a TI, it breaks time-reversal symmetry. This opens a gap in the SS and gives rise to dissipationless quantized edge transport \cite{Toku2019,He2018,Liu2016,Wang2015,Qi2008,Mong2010,Chang2013}. 
Therefore, magnetic TIs (MTIs) can host a set of emergent phenomena such as quantum anomalous Hall effect, axion insulating state, and quantum magnetoelectric effect \cite{Toku2019,He2018,Liu2016,Wang2015,Qi2008,Mong2010,Chang2013,Li2019}. 
Similarly, engineering SC in topological materials is generally believed to be an effective approach to achieving topological SC (TSC), which can support Majorana fermions at the edge states and Majorana zero mode at vortex cores, which are potentially useful in topological quantum computing \cite{Elli2015,Qi2009,Fu2008,Schn2008,Ryu2010,Akhm2009,Wilc2009,Naya2008,Wang2018}. 
Despite theoretical advances indicating that over half of known materials may possess nontrivial band topology \cite{Brad2017,Verg2019,Verg2022}, only a very limited number of materials exhibit either intrinsic magnetism or SC. This scarcity hinders the search for intrinsic MTI or TSC. 
The exploration of new magntic or superconductive materials with topological band structures remains a prominent and challenging frontier in condensed matter physics.

The layered pseudobinary chalcogenides $mAX\cdot nB_2X_3$ ($A$ = Ge, Sn, Pb, Mn; $B$ = Sb, Bi; $X$ = Se, Te; $m$ = 0, $n$ = 1 or $m$ = 1, $n \geq$ 1 or $m \geq$ 2, $n$ = 1) have attracted extensive research interest due to their nontrivial band topology which host exotic topological physical properties \cite{Nurm2020,Okam2012,Neup2012,Li2021,Nies2014,Frag2021,Soum2012,Kuro2012,Erem2012,Okud2013,Papa2016,Paci2018,Zhang2019,Li2019,Otro2019,Gong2019,Hao2019,Chen2019,Li2019b,Sa2012,Kim2012,Li2023,Gao2023}. 
Moreover, SC has also been widely explored in this material family.
For example, pressurized Bi$_2$Se$_3$ and Bi$_2$Te$_3$ become bulk superconductors with transition temperature $T_c$ up to $\sim$ 7 K \cite{Kirs2013,Zhang2011}. Chemical pressure induced by In doping in SnBi$_2$Te$_4$ and PbBi$_2$Te$_4$ also leads to SC near 2 K \cite{McGu2023,Xu2023}. 
Moreover, several nonmagnetic members—including GeBi$_2$Te$_4$, GeBi$_4$Te$_7$, GeSb$_2$Te$_4$, SnBi$_2$Te$_4$, and so on—exhibit pressure-induced SC with maximum $T_c$ values ranging from 6.3 to 8.5 K \cite{Liu2024,Huang2025,Gree2017,Zhou2023,Li2022,Song2020}.
In contrast, among magnetic members, SC remains scarce, with MnSb$_4$Te$_7$ being the only example reported so far, showing a low $T_c$ of about 2 K at 50 GPa \cite{Pei2022}.

Theoretical predicts that Ge$_2$Bi$_2$Te$_5$ is TI with a single Dirac cone \cite{Kim2012}, which has been confirmed experimentally by our recent work \cite{Tian2025}.
On the other hand, recent study has reported the successful Mn doping into Ge$_2$Bi$_2$Te$_5$, although the synthesis of pure Mn$_2$Bi$_2$Te$_5$ remains challenging \cite{Qian2024}. 
(Ge$_{1-x}$Mn$_x$)$_{2}$Bi$_2$Te$_5$ single crystals with $x$ up to 0.47 can be grown and the $x$ = 0.47 sample shows an antiferromagnetism (AFM) with $T_{\rm N}\sim $ 11 K \cite{Qian2024}.
In this work, we carry out a systematic study on electrical transport properties of (Ge$_{1-x}$Mn$_x$)$_{2}$Bi$_2$Te$_5$ (0 $\leq x \leq$ 0.49) under pressure. 
We discover SC in Ge$_2$Bi$_2$Te$_5$ with maximum $T_c\sim$ 7.6 K at 23.0 GPa. Furthermore, it is found that Mn doping induces AFM at ambient pressure but suppresses SC quickly under high pressure, implying the competing relationship between AFM and SC in this material series.

\section{Methods}

\textbf{Sample synthesis.} 
Single crystals of Ge$_2$Bi$_2$Te$_5$ were grown by the self-flux method. The high-purity Ge (99.999 \%), Bi (99.999 \%) and Te (99.999 \%) shots were put into corundum crucibles and sealed into quartz tubes with a ratio of Ge : Bi : Te = 2 : 2 : 8. The tube was heated to 1273 K at a rate of 40 K/h and held there for 12 h to ensure a homogeneous melt. Then the temperature was rapidly cooled down to 672 K with subsequently cooling down to 593 K at 1 K/h. The flux is removed by centrifugation, and shiny crystals can be obtained. 
For Mn-doped single crystals (Ge$_{1-x}$Mn$_x$)$_2$Bi$_2$Te$_5$, they are difficult to synthesize by using the flux method, and the chemical vapour transport (CVT) method needs to be used \cite{Qian2024}. Ge (99.999 \%), Mn (99.9 \%), Bi (99.999 \%) and Te (99.999 \%) powders with the molar ratio of Ge/Mn : Bi : Te = 4 : 2 : 6 were put in a silica tube with a length of 200 mm and an inner diameter of 15 mm. Then, 200 mg I$_2$ was added into the tube as a transport reagent. The tube was evacuated to 10$^{-2}$ Pa and sealed under vacuum. The tubes were placed in a two-zone horizontal tube furnace, slowly heated to 1273 K, and held at this temperature for 24 h. Then the temperatures of the source and growth zones were cooled to 813 K and 793 K, respectively, and held there for two weeks. Shiny crystals with lateral dimensions of up to several millimetres can be obtained.

\textbf{Structural and composition characterizations.} 
The X-ray diffraction (XRD) patterns of (Ge$_{1-x}$Mn$_x$)$_2$Bi$_2$Te$_5$ single crystals were performed using a Bruker D8 X-ray diffractometer with Cu $K_{\alpha}$ radiation ($\lambda$ = 1.5418 \AA) at room temperature. The elemental analysis was performed using energy-dispersive X-ray (EDX) spectroscopy analysis in a FEI Nano 450 scanning electron microscope.

\textbf{Transport and magnetization characterizations.}
Electrical transport measurements were conducted using a superconducting magnetic system (Cryomagnetics, C-Mag Vari-9). Magnetization measurements were performed using a Quantum Design magnetic property measurement system (MPMS3).

\textbf{High Pressure Transport.}
\textit{In--situ} high-pressure electrical transport measurements were carried out in the Quantum Design Physical Property Measurement System (PPMS-9T). The van der Pauw method was employed to measure the resistivity. Small pieces of (Ge$_{1-x}$Mn$_{x}$)$_{2}$Bi$_{2}$Te$_{5}$ single crystal were loaded into the chamber of diamond anvil cell (DAC) with 200 $\mu$m-sized diamond culets. The ruby luminescence at room temperature was utilized to obtain the pressure values within the DAC \cite{Mao1986}. 
	
\section{Results and Discussion}

As depicted in Fig. \ref{FigStr}(a), (Ge$_{1-x}$Mn$_x$)Bi$_2$Te$_5$ crystallizes in a layered rhombohedral structure with space group $P\bar{3}m1$ (No. 164) \cite{Shel2000, Karp2000, Alak2021,Qian2024}, where the fundamental building block is a nonuple layer (NL) with the stacking sequence Te–Bi–Te–Ge/Mn–Te–Ge/Mn–Te–Bi–Te, and these NLs are stacked along the $c$-axis via weak vdW interactions. 
The EDX spectra show that the actual compositions of present crystals are Ge$_{1.91(2)}$Bi$_{2.11(2)}$Te$_5$ (noted as $x$ = 0), (Ge$_{0.75}$Mn$_{0.25}$)$_{1.77(2)}$Bi$_{2.18(2)}$Te$_5$ ($x$ = 0.25), and (Ge$_{0.51}$Mn$_{0.49}$)$_{1.81(2)}$Bi$_{2.17(2)}$Te$_5$ ($x$ = 0.49) when setting Te content as 5.
The deviation from ideal stoichiometry of 2 : 2 : 5 suggests that there are some Ge$_{\rm Bi}$ antisite defects in this series of materials, consistent with previous study \cite{Qian2024}. 
Single-crystal XRD patterns at ambient pressure (Fig. \ref{FigStr}(b)) show that all reflections can be indexed to the (00$l$) peaks of the trigonal $P\bar{3}m1$ space group, confirming that the crystallographic $c$-axis is perpendicular to the sample surface. 
Moreover, with increasing Mn content $x$, the (00$l$) peaks shift systematically to higher angles (left inset of Fig. 1(b)), reflecting a gradual decrease in the $c$-axial lattice parameter from 17.37(1) \AA\ ($x$ = 0) to 17.29(1) \AA\ ($x$ = 0.25) and 17.24(1) \AA\ ($x$ = 0.49) (right inset of Fig. 1(b)), consistent with the smaller size of Mn ion than that of Ge ion. 
The temperature-dependent $c$-axial magnetic susceptibility $\chi_{c}(T)$ (Figs. \ref{FigStr}(c) and S1(a) in the Supplemental Material \cite{SM}) reveals antiferromagnetic (AFM) order in the Mn-doped samples. The Néel temperature $T_\text{N}$ increases gradually from 6.0 K to 11.8 K when the $x$ increases from 0.25 to 0.49, in agreement with the results in the literature \cite{Qian2024}. 
Figures \ref{FigStr}(d) and S1(b) \cite{SM} shows the temperature dependence of in-plane normalized resistivity $\rho_{xx}(T)/\rho_{xx}$(2 K). All of (Ge$_{1-x}$Mn$_x$)Bi$_2$Te$_5$ crystals show metallic behaviors, i.e., the $\rho_{xx}(T)$ decreases with cooling from 300 K down to 2 K in general. 
It is noted that a kink appears in the Mn-doped samples at low-temperature region and the corresponding temperatures are close to the $T_\text{N}$s, which can be ascribed to the suppression of spin scattering upon entering the AFM state.

\begin{figure}
	\includegraphics[scale=0.18]{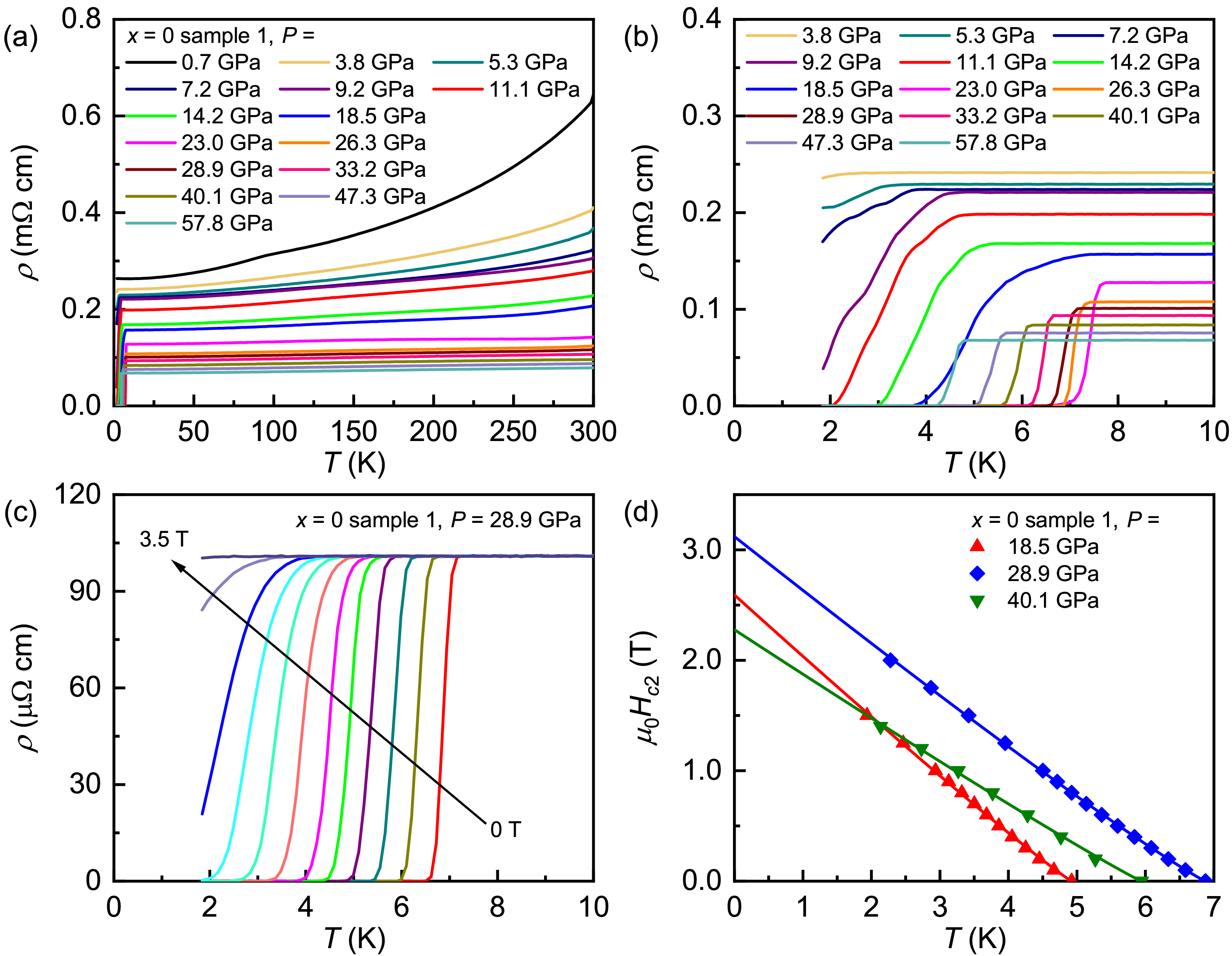}
	\caption{
		(a) Temperature dependence of resistivity $\rho(T)$ at pressures ranging from 0.7 to 57.8 GPa in Ge$_{2}$Bi$_{2}$Te$_{5}$ ($x$ = 0, sample 1). 
		(b) The enlarged $\rho(T)$ curves below 10 K. 
		(c) The $\rho(T)$ as a function of temperature under various magnetic fields at 28.9 GPa. 
		(d) The temperature dependence of $\mu_{0}H_{c2}(T)$ at 18.5, 28.9 and 40.1 GPa, respectively. The solid lines represent the fitting results using the equation $\mu_{0} H_{c2}(T) = \mu_{0} H_{c2}(0)(1-(T/T_{c}))^{1+\alpha}$.
	}
	\label{Fighp1}
\end{figure}

Figures \ref{Fighp1}(a) and \ref{Fighp1}(b) show the high-pressure transport results of the Ge$_2$Bi$_2$Te$_5$ single crystal.
At 0.7 GPa, the $\rho(T)$ curve shows a metallic behavior in the whole temperature range, which is in line with that under ambient pressure. 
With increasing pressure, the values of  $\rho(T)$ decrease gradually. Importantly, there is a slight drop in $\rho(T)$ curve at the onset temperature ${T}_{c,\text{onset}} \sim$ 2.46 K under a pressure of 3.8 GPa.
As the pressure increases to 11.1 GPa, the $\rho(T)$ curve drops significantly at ${T}_{c,\text{onset}} \sim$ 3.99 K and reaches zero at ${T}_{c,\text{zero}} \sim$ 1.96 K. 
It suggests that such resistivity drop is due to the pressure-induced SC in Ge$_2$Bi$_2$Te$_5$. 
Upon further compression, the ${T}_{c,\text{onset}}$ initially increases and approaches the maximum value of $\sim$ 7.58 K at 23.0 GPa. 
Subsequently, the ${T}_{c,\text{onset}}$ begins to decrease monotonically under higher pressure. Furthermore, the repeated measurements conducted on another Ge$_{2}$Bi$_{2}$Te$_{5}$ single crystal also exhibit the similar superconducting phenomena, as depicted in Figs. S2(a) and S2(b) \cite{SM}. 
To gain further insight into the superconducting state, the $\rho(T)$ curves under different magnetic fields at representative pressures were measured (Figs. 2(c), S3(a) and S3(b) \cite{SM}). 
For $P$ = 28.9 GPa (Fig. 2(c)), the ${T}_{c,\text{onset}}$ is gradually suppressed  with increasing field and it can not be observed above 2 K when $\mu_{\rm 0} H$ = 3.5 T.
Using the criterion of 50 \% normal state resistivity just above ${T}_{c,\text{onset}}$, the temperature dependence of upper critical field $\mu_{0}H_{c2}(T)$ at selected pressures can be extracted, as shown in Fig. 2(d). 
The $\mu_{0}H_{c2}(T) = \mu_{0}H_{c2}(0)(1-(T/T_{c}))^{1+\alpha}$ equation \cite{Mull2001} is used to fit these curves, where $\mu_{0} H_{c2}(0)$ represents the upper critical field at 0 K. 
The fitted value of $\mu_{0}H_{c2}(0)$ are 2.59(3) T for 18.5 GPa, 3.12(2) T for 28.9 GPa and 2.28(5) T for 40.1 GPa, respectively.
As the Pauli limiting field $\mu_{0}H_{\rm P}(0)$ (= 1.86 $T_{c}$) $\sim$ 9.15 T -- 11.05 T, which are larger than above values of $\mu_{0}H_{c2}(0)$, it suggests that the orbital-depairing mechanism should be dominant.

\begin{figure}
	\includegraphics[scale=0.18]{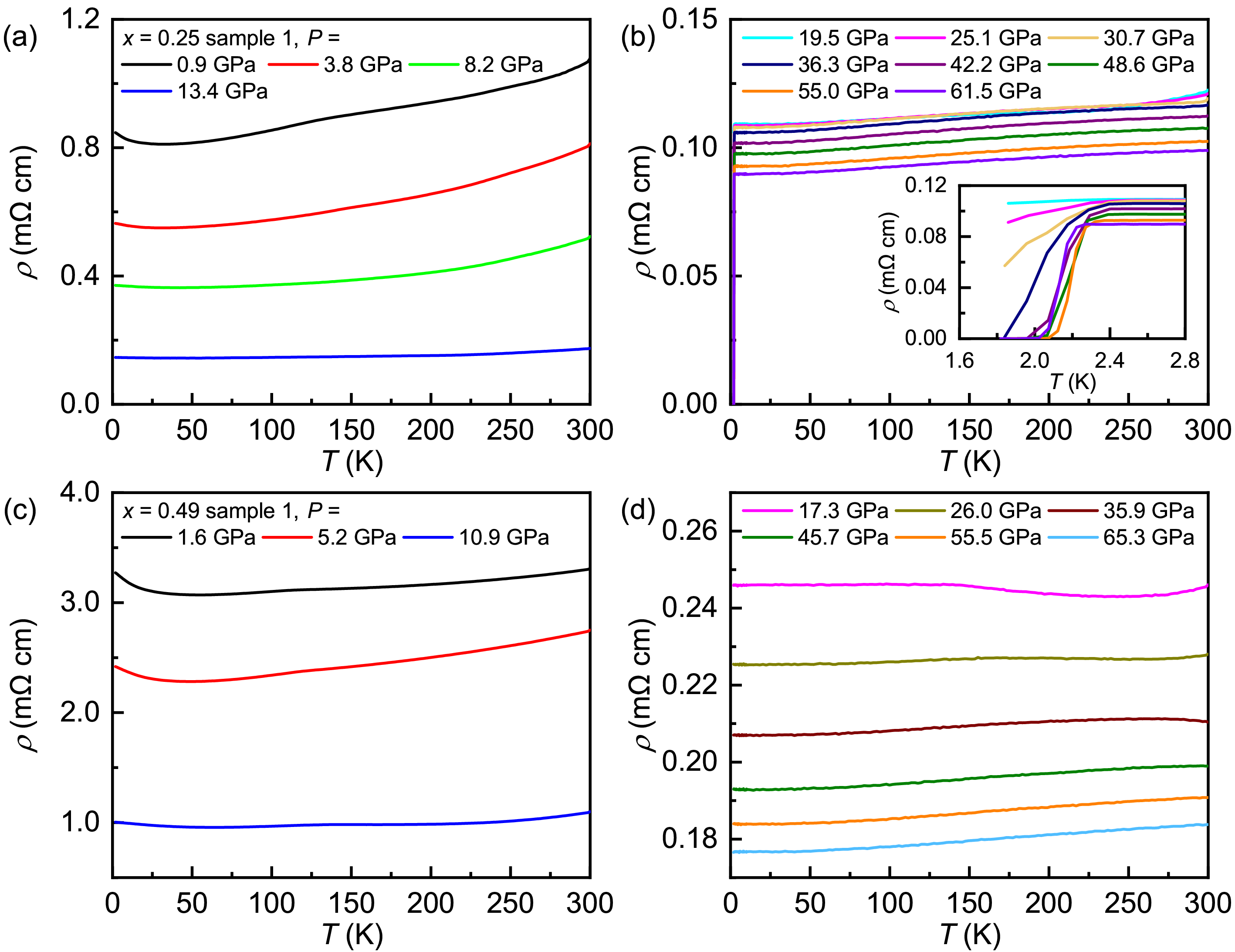}
	\caption{
		(a) and (b) Temperature dependence of $\rho(T)$ under pressures up to 61.5 GPa for $x$ = 0.25 sample. The inset of (b) shows the enlarged $\rho(T)$ curves from 1.8 to 4 K.
		(c) and (d) The $\rho(T)$ curves in the pressure range from 1.6 to 65.3 GPa for $x$ = 0.49 sample.
	}
	\label{Fighp2}
\end{figure}

We now turn to the pressure effects on the Mn doped Ge$_{2}$Bi$_{2}$Te$_{5}$ single crystals.
For $x$ = 0.25 sample (Figs. \ref{Fighp2}(a) and \ref{Fighp2}(b)), the values of $\rho(T)$ initially decrease dramatically as pressure increases. Then the $\rho(T)$ curve exhibits a metallic behavior in the whole temperature range at the pressure of 19.5 GPa. 
Simultaneously, the $\rho(T)$ curve drops slightly at $T_{c, \rm onset}$ $\sim$ 2.3 K. 
Upon further compression, the values of normal-state $\rho(T)$ decline slowly and the drop becomes more dramatic while $T_{c, \rm onset}$ remains almost unchanged. 
When the pressure is applied at 42.2 GPa, $T_{c, \rm onset}$ is about 2.26 K and the $T_{c, \rm zero}$ is detected at approximately 1.84 K. 
When the pressure is further increased to 61.5 GPa, both $T_{c, \rm onset}$ and $T_{c, \rm zero}$ present slight variations, reaching $\sim$ 2.20 K and 1.97 K at 61.5 GPa, respectively. 
Furthermore, the temperature dependence of $\rho(T)$ under various magnetic fields at a pressure of 55.0 GPa is depicted in Fig. S4(a) \cite{SM}. 
It can be seen that the SC is suppressed quickly by applying only a small field of 0.3 T, which is much smaller than that in undoped Ge$_{2}$Bi$_{2}$Te$_{5}$. 
The $\mu_{0}H_{c2}(T)$ curve can also be described by the equation $\mu_{0} H_{c2}(T) = \mu_{0} H_{c2}(0)(1-(T/T_{c}))^{1+\alpha}$, and the fitted $\mu_{0} H_{\rm c2}(0)$ is 1.14(6) T, confirming the dramatically decrease of $\mu_{0} H_{c2}(0)$ with Mn doping. 
By contrast, the highly doped sample with $x$ = 0.49 also exhibits the analogous resistivity behavior under high pressure, as illustrated in Figs. \ref{Fighp2}(c) and \ref{Fighp2}(d).
The values of $\rho(T)$ is suppressed by one order of magnitude when the pressure approaches 65.3 GPa. Nevertheless, distinct from the samples with $x$ = 0 and 0.25, no superconducting transition appears in our measured pressure range. 
Additionally, the  transport behaviors under pressure of $x$ = 0.25 and 0.49 samples can be reproduced, as presented in Figs. S5 and S6 \cite{SM}. 

\begin{figure}
	\includegraphics[scale=0.25]{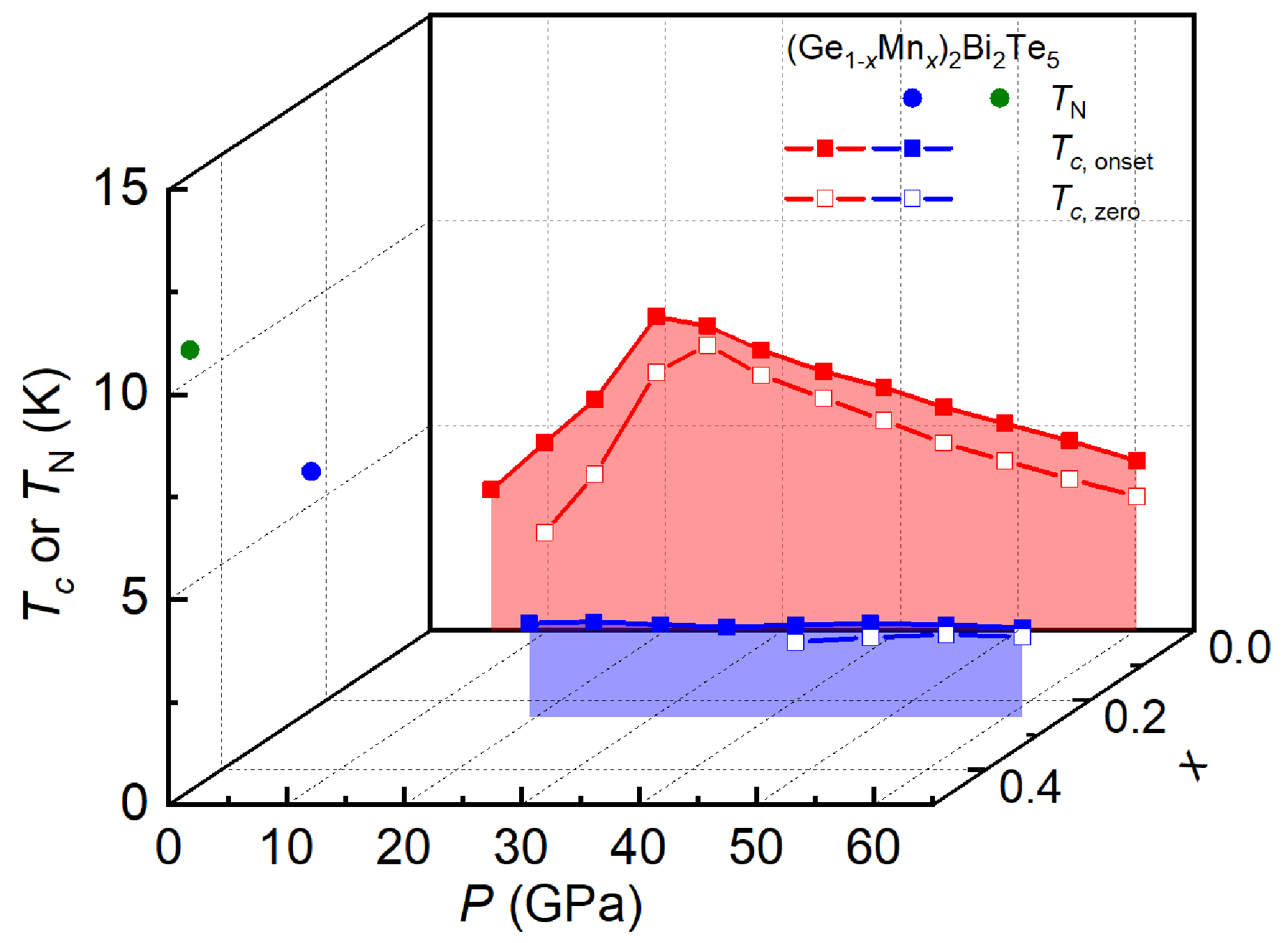}
	\caption{
		Electronic phase diagram of (Ge$_{1-x}$Mn$_{x}$)$_2$Bi$_2$Te$_5$ ($x$ = 0, 0.25 and 0.49).
	}
	\label{FigPha}
\end{figure}

We summarized the electronic phase diagram of (Ge$_{1-x}$Mn$_x$)$_2$Bi$_2$Te$_5$ with the variations of pressure and $x$ (Fig. \ref{FigPha}). 
For Ge$_2$Bi$_2$Te$_5$, which is paramagnetic at ambient pressure, the pressure dependence of $T_c$ shows a dome-shaped feature. 
SC emerges above $\sim$ 4 GPa, boosts to the maximum value of 7.6 K at 23.0 GPa and then decreases with increasing pressure further. 
For $x$ = 0.25 sample with $T_{\rm N}$ $\sim$ 6 K, superconducting region is significantly narrowed. The pressure exceeding 19 GPa is required to induce SC, and the $T_c$ forms a plateau at approximately 2.3 K over a wide pressure range from 20 to 60 GPa. 
In the case of higher Mn doped sample ($x$ = 0.49) with higher $T_\text{N}$ $\sim$ 12 K, SC is further suppressed, which can not be observed up to the maximum pressure of 65 GPa.
These results suggest that AFM order and SC are competitive each other, indicating that magnetic interaction may have adverse effects on the formation of Cooper pairs in (Ge$_{1-x}$Mn$_x$)$_2$Bi$_2$Te$_5$.

\begin{table*}
	\caption{
		SC induced by pressure or chemical doping in pseudobinary chalcogenides $mAX\cdot nB_2X_3$ ($A$ = Ge, Sn, Pb, Mn; $B$ = Sb, Bi; $X$ = Se, Te), along with their maximum $T_c$ values. Here, ``$\times$'' indicates no SC observed above 2 K, while ``/'' denotes no relevant reports.}
	\begin{ruledtabular}
		\begin{tabular}{llllll}
			Nonmagentic & Maximum $T_c$ & Reference & Magnetic & Maximum $T_c$  & Reference \\
			\hline
			Ge$_2$Bi$_2$Te$_5$ & $\sim$ 7.6 K at 23.0 GPa & This work & (Ge$_{1-x}$Mn$_{x}$)$_2$Bi$_2$Te$_5$ ($x$ = 0.25) & 2.3 K at 55.0 GPa & This work \\
			& & & (Ge$_{1-x}$Mn$_{x}$)$_2$Bi$_2$Te$_5$ ($x$ = 0.49) & $\times$ & This work \\
			GeBi$_2$Te$_4$ & $\sim$ 8.4 K at 14.6 GPa & \cite{Liu2024} & MnBi$_2$Te$_4$ & $\times$ & \cite{Pei2020} \\
			GeBi$_4$Te$_7$ & $\sim$ 8.3 K at 20.0 GPa & \cite{Huang2024} & MnBi$_4$Te$_7$ & $\times$ & \cite{Pei2020,Shao2021} \\
			GeSb$_2$Te$_4$ & $\sim$ 8 K at 22 GPa & \cite{Gree2017} & MnSb$_2$Te$_4$ & $\times$ & \cite{Yin2021} \\
			GeSb$_4$Te$_7$ & $\sim$ 8 K at 34 GPa & \cite{Zhou2023} & MnSb$_4$Te$_7$ & $\sim$ 2.2 K at 50.7 GPa & \cite{Pei2022} \\
			SnBi$_2$Te$_4$ & $\sim$ 8.5 K at 22 GPa & \cite{Li2022} & MnBi$_6$Te$_{10}$ & $\times$ & \cite{Shao2021} \\
			SnSb$_2$Te$_4$ & $\sim$ 7.4 K at 33 GPa & \cite{Song2020} & MnSb$_6$Te$_{10}$ & / & \\
			Bi$_2$Se$_3$ & $\sim$ 7 K at 30 GPa & \cite{Kirs2013} & & & \\
			Bi$_2$Te$_3$ & $\sim$ 8 K at 9 GPa & \cite{Zhang2011} & & & \\
			Sb$_2$Te$_3$ & $\sim$ 6.3 K at 7.5 GPa & \cite{Zhu2013} & & & \\
			BiTe & $\sim$ 8.4 K at 11.9 GPa & \cite{Zhu2024} & & & \\
			Sn$_{1-x}$In$_x$Bi$_2$Te$_4$ & $\sim$ 1.8 K for $x$ = 0.6 & \cite{McGu2023} & & & \\
			Pb$_{1-x}$In$_x$Bi$_2$Te$_4$ & $\sim$ 2.1 K for $x$ = 0.7 & \cite{Xu2023}& & & 
		\end{tabular}
	\end{ruledtabular}
	\label{Tabsc}
\end{table*}

Finally, we summarize the pressure-induced SC in (Ge$_{1-x}$Mn$_x$)$_2$Bi$_2$Te$_5$ and other members of the $mAX\cdot nB_2X_3$ family in Table \ref{Tabsc}, including both magnetic and non-magnetic systems. 
Among the non-magnetic members, those $B_2X_3$ and $AX\cdot nB_2X_3$ phases exhibit SC under pressure. Notably, we report here the first observation of pressure-induced SC among $2AX\cdot B_2X_3$-type materials. 
Strikingly, the maximum $T_c$ across these systems are very similar, falling in the range of 6.3 -- 8.5 K, although the required pressures vary considerably, from about 8 to 34 GPa. 
This suggests that pressure-induced SC in the $mAX\cdot nB_2X_3$ family may share a common underlying origin or mechanism. 
It is noted that In-doped SnBi$_2$Te$_4$ and PbBi$_2$Te$_4$ exhibit SC even at ambient pressure although the $T_c$ is rather low ($\sim$ 2 K).
It would be interesting to explore the evolution of SC under pressure in these systems.
As for magnetic members, MnSb$_4$Te$_7$ is the only one hosting pressure-induced SC reported in the literature. 
Our present work establishes that applying pressure to Ge$_2$Bi$_2$Te$_5$ with low Mn-doping levels can trigger a transition from an AFM ground state at ambient pressure to a superconducting state, thereby adding a new magnetic member to this category. 
Moreover, the $T_c$ of $x$ = 0.25 sample is only about 2 K, similar to that of MnSb$_4$Te$_7$, and both of them have comparable $T_{\rm N}$s. It implies that there is a competing relationship between AFM and SC, and magnetic interactions in these systems may have a similar depairing effect on SC.

\section{Conclusion}

In summary, we discovered the pressure-induced SC in non-magnetic TI Ge$_2$Bi$_2$Te$_5$ with its $T_c$ following a dome-shaped evolution as a function of pressure. 
Furthermore, the introduction of Mn doping leads to the emergence of AFM order at ambient presurre and strongly suppresses the $T_c$ of pressure-induced SC. 
Similar behaviors have been observed in GeSb$_4$Te$_7$ and MnSb$_4$Te$_7$. It suggests that within this family of topological materials, magnetic interactions appear to be detrimental to the formation of SC. 
Present study provides experimental insights for a deeper understanding of the relationship between band topology, magnetism, and SC in $mAX\cdot nB_2X_3$ material family. It also offers a new material platform for exploring TSC and correlated topological states.

$^{\dag}$ These authors contributed equally to this work.

$^{\ast}$ Corresponding authors: X. Z. (zhangxiaobupt@bupt.edu.cn), Y. P. Q. (qiyp@shanghaitech.edu.cn), H. C. L. (hlei@ruc.edu.cn), S. G. W. (sgwang@ahu.edu.cn)

\begin{acknowledgments}
This work was supported by the National Key R\&D Program of China (Grants No. 2022YFA1403800, 2023YFA1406500 and 2023YFA1607400), the National Natural Science Foundation of China (Grants No. 12274459, 52130103 and 52572288), and the China Postdoctoral Science Foundation (Grant No. 2023M730011).
\end{acknowledgments}


%

\end{document}